# FINITE STATE MACHINE BASED EVALUATION MODEL FOR WEB SERVICE RELIABILITY ANALYSIS

Thirumaran.M[1], Dhavachelvan.P[2], S.Abarna[3] and Lakshmi.P[4]

[1,3,4] Department of Computer Science and Engineering, Pondicherry Engineering College, India.

[1]thirumaran@pec.edu

[3]abarna.pec@gmail.com

[4]lakshmip2990@gmail.com

[2]Department of Computer Science and Engineering, Pondicherry University, India.

[2]dhavachelvan@gmail.com

## *ABSTRACT*

*Today's world economy demands that both market access and customer service be available anytime and anywhere. The Web is the only way to supply global economic needs and, due to expand the development of comprehensive web service, it does so relatively inexpensively. The ability of web service is to provide a relatively inexpensive way to deploy customer services. As days goes on the business logic of a system emerges out at a great extent where it has to react to several different competitors under different situations. Through means of a business logic system we can able to achieve faster communication of information, rampant change and increasing business complexity. As competitors gets increased, so it causes the business parties to improve by promoting the business process management to the standard level by incorporating the higher-end technology solution such as service oriented business communication and business rule management automation. Now-a-days they are very much considering about the changes to be done at shorter time since the reaction time needs are decreasing every moment. Business Logic Evaluation Model (BLEM) are the proposed solution targeting business logic automation and facilitating business experts to write sophisticated business rules and complex calculations without costly custom programming. BLEM is powerful enough to handle service manageability issues by analyzing and evaluating the computability and traceability and other criteria of modified business logic at run time. The web service and QOS grows expensively based on the reliability of the service. Hence the service provider of today things that reliability is the major factor and any problem in the reliability of the service should overcome then and there in order to achieve the expected level of reliability. In our paper we propose business logic evaluation model for web service reliability analysis using Finite State Machine (FSM) where FSM will be extended to analyze the reliability of composed set of service i.e., services under composition, by analyzing reliability of each participating service of composition with its functional work flow process. FSM is exploited to measure the quality parameters. If any change occurs in the business logic the FSM will automatically measure the reliability.*

# 1. INTRODUCTION

The Business Logic Management system (BLMS) brings about best results if we select a BLMS that is able to reconcile the business methodology of that system. The BLMS has the following features which makes it a better system. BLMS is fully scalable, it can be integrated, and it separates the business logic from the clients who deplete that information. This result in intense levels of code reuse and portability as well as it enables applications to be upgraded without affecting testing and functioning of business logic. It allows the non- technical people to create and alter very detailed levels of business logic without calling for the assistance of the IT Department, by providing a point-and-click or GUI interface that endows the user to manage the logic using easily apprehensible symbols. The system supports a test environment that permits the business analyst or non technical user to test logic changes before deploying the modifications in a live environment. This is the only means to ascertain that changes do not negatively impact business logic integrity or compromise the production system. In support of this test environment, the BLMS has a realistic debugging process that is graphical and easy to understand. The debugger provides a method for adjusting parameters and data, in real time, so that the business analyst can create realistic test cases and validate each one as they are executed.

The BLMS is modular. Each module can be easily integrated with existing modules so that the applications can acquire added functionality as the need arises. Those modules provide built-in support for the core requirements of any advanced business system which includes the ability to automate decision making using a rule-based process, a monitoring engine that lets us configure any number of check points, support for integral workflow and BPM (Business Process Management) capabilities and the ability to communicate with other systems.

BLMS transparently communicates with the information consumers want without forcing those consumers to know anything about the underlying technology of the BLMS including the language it was written in, the version of the compiler used, or the type and structure of any associated databases. In essence, the BLMS is capable of putting up a "black box" interface where the consumer simply sends a query and receives the data back in a format that it can interpret. It is capable of supporting web services and other current technologies like HTML, XML, SOAP, etc. The IT Department does not have to hold out an excessive learning when implementing the BLMS. They have to build a complete documentation and actual code examples to make the transition to a logic-based system easier and to accelerate the implementation and deployment process. It includes accession to a full SDK (Software Developers Kit) that enables developers to easily create and integrate their own custom components into the BLMS without negatively impacting the system's usability, security or stability. Thus, the goal of implementing BLMS is to allow the organization to ultimately reduce the cost by automating tasks that used to require large human intervention. Let us have a detailed look at the essential component of a BLMS namely the Business Rule Management System and the Business Logic Evaluation Model for the automation of services.

# 2. LITERATURE SURVEY

Web services used primarily as a means for businesses to communicate with each other and with clients, Web services allow organizations to communicate data without intimate knowledge of each other's IT systems behind the firewall. Unlike traditional client/server models, such as a Web server/Web page system, Web services do not provide the user with a GUI. Web services instead share business logic, data and processes through a programmatic interface across a network. The applications interface, not the users. Developers can then add the Web service to a GUI (such as a Web page or an executable program) to offer specific functionality to users.

Web services allow different applications from different sources to communicate with each other without time-consuming custom coding, and because all communication is in XML, Web services are not tied to any one operating system or programming language. For example, Java can talk with Perl, Windows applications can talk with UNIX applications. This is made possible by using technologies such as Jini, UPnP, SLP, etc.

Slim Trabelsi and Yves Roudier proposed a scalable solution to enabling secure and decentralized discovery protocols. It also deals how to extend the WS-Discovery Web Service protocol with these mechanisms [1]. Colin Atkinson and Philipp Bostan proposed the brokerage aspect of the web service vision but it is difficult to involve in setting up and maintaining useful repositories of web services. So they describe a pragmatic approach to web service brokerage based on automated indexing and discuss the required technological foundations [2].

Janette Hicks and Weiyi Meng proposed a current discovery research through use of the Google Web service, UDDI category searching and private registry. They found WSDL documents for a given domain name, parse the desired service document to obtain invocation formats, and automatically invoke the Web service to support enhancements of HTML-dependent search tools by providing access to data inaccessible through surface HTML interfaces [3].

ZHANG Changyou and ZHU Dongfeng invented a web service discovery mechanism on unstructured P2P network. The web services are clustered into communities through functional properties and several query packets will be proliferated and spread through the community. Each service in this community will be evaluated through non-functional properties. The service clustering and experience exchanging enhanced the efficiency in discovery [4]. Henry Song and Doreen Cheng examine better approaches of using general-purpose search engines to discover Web Services. They used Yahoo and Google search engine and the queries were fired to each search engine daily and the top 100 search results returned from every search are collected and analyzed. The results show that for both search engines, embedding a WSDL specification in a Web page that provides semantic description of the service [5].

A QoS-oriented Optimization Model for Web Service Group proposed by Xiaopeng Deng and Chunxiao Xing defines functionality satisfaction degree, performance satisfaction degree, cost satisfaction degree and trust satisfaction degree as the QoS parameters of a Web service. The Web service optimization model is described formally. simulation cases prove that the proposed model is effective and can be used for Web service selection from its group [6]. Zibin Zheng, Hao Ma, Michael R. Lyu, and Irwin King came up with a Collaborative Filtering Based Web Service Recommender System which includes a user-contribution mechanism for Web service QoS information collection and an effective and novel hybrid collaborative filtering algorithm for Web service QoS value prediction. The comprehensive experimental analysis shows that WSRec achieves better prediction accuracy than other approaches [7].

Hao Yang, Junliang Chen, Xiangwu Meng and Ying Zhang put forth a Dynamic Agent based Web Service Invocation Infrastructure which presents a Web service invocation infrastructure based on software agents [8]. Youngkon Lee set forth Web Services Registry implementation for Processing Quality of Service. He proposed the design principle for integrating quality management on Web service registry developed in UDDI specification and Web service quality management system (WSQMS). For representing Web service quality information, the WSQDL (Web Service Quality Description Language), which published by WSQM technical committee in OASIS is adopted. In more detail, this paper also presents the scheme to compose the classification scheme for quality data and to modify the necessary data structure of the registry [9]. Pat. P. W. Chan and Michael R. Lyu proposed Dynamic Web Service Composition which is a new approach in building reliable web service. In this paper, a dynamic web service composition algorithm with verification of Petri-Nets is provided. A series of experiments were conducted to evaluate the correctness and performance of the composed web service [10].

The paper by Haibo Zhao and Prashant Doshi discusses Automated RESTful Web Service Composition. A formal model for describing individual Web services and automating the composition has been proposed. The paper represents the initial efforts towards the problem of automated RESTful Web service composition [11].

The paper Signature-Based Composition of Web Services deals with the problem of querying a UDDI registry with a functional specification of a service, and getting in return a single service, or a composition of services that address the functional need. The approach behind relies on service signatures (message types). The principles, algorithms, implementation and results are discussed and proposed by Aniss Alkamari, Hafedh Mili and Abdel Obaid [12].

Composition and Reduction of Web Service Based on Dynamic Timed Colored Petri Nets introduced by Yaojun Han and Xuemei Luo presents the formalization of Web services and the algorithm for constructing composition. The Dynamic Timed Colored Petri Net (DTCPN) is used to model the dynamic behavior of large and complex systems. A reduction algorithm of DTCPN for the basic structures of the Web service composition is put forth. The correctness and time and cost performance of the Web service composition is also discussed [13]. Wen-Yau Liang came up with Apply Rough Set Theory into the Web Services Composition which develops a generic genetic algorithm incorporating knowledge extracted from the rough set theory. It includes improving the performance of the Genetic Algorithms (GA) by reducing the domain range of initial population, rule constraining crossover process and rule constrained mutation process, using the rough set theory for composite web services producing an optimal solution [14].

An Approach to Composing Web Services with Context Heterogeneity is set forth by Xitong Li, Stuart Madnick, Hongwei Zhu, and Yushun Fan. The contexts of the involved Web services are defined in a lightweight ontology and their WSDL descriptions are annotated by an extension of a W3C standard and its composition is describe using BPEL specification that ignores context heterogeneity and it automatically detects all context differences among the involved services, and reconciles them by producing a mediated BPEL file with necessary conversions using XPath functions and/or Web services [15]. An infrastructure that allows users and applications to discover, deploy, compose and synthesize services automatically is needed for web services. An approach for automatic service composition based on semantic description of Web services is implemented and will be used for the WS-Challenge 2007. The above said idea is put forward in the paper Semantics-Based Web Service Composition Engine by Srividya Kona, Ajay Bansal, Gopal Gupta and Thomas D. Hite [16].

A Global QoS Optimizing Web Services Selection Algorithm based on MOACO for Dynamic Web Service Composition has been proposed by Fang Qiqing, Peng Xiaoming, Liu Qinghua and Hu Yahui. The paper presents a novel global QoS optimizing and multi-objective Web Services selection algorithm based on Multi-objective Ant Colony Optimization (MOACO) for the Dynamic Web Service composition. They have proposed a model that converts the mono-objective problem into a multi-objective optimization problem with user constraints. The experimental results prove that the proposed algorithm outperforms the recently published QoS Global Optimization Based on Multi-objective Genetic Algorithm (MOGA) [17].

The paper A Reflective Framework to Improve the Adaptability of BPEL-based Web Service Composition presented by Yanlong Zhai, Hongyi Su and Shouyi Zhan presents the reflective framework here aims to improve the adaptability of BPEL-based web service composition. A meta-model was defined to build the self-representation of the web services composition. A prototype adaptive service composition environment has been developed to implement the reflective framework and demonstrate its effectiveness on providing adaptive web services composition. [18]. San-Yih Hwang came up with the Dynamic Web Service Selection for Reliable Web Service Composition which aims to determine a subset of Web services to be invoked at runtime so as to successfully orchestrate a composite Web service. A finite state machine to model the permitted invocation sequences of Web service operations has been devised. This paper also discusses the strategies to select Web services that are likely to successfully complete the execution of a given sequence of operations [19].

## 3. BUSINESS LOGIC EVALUATION MODEL

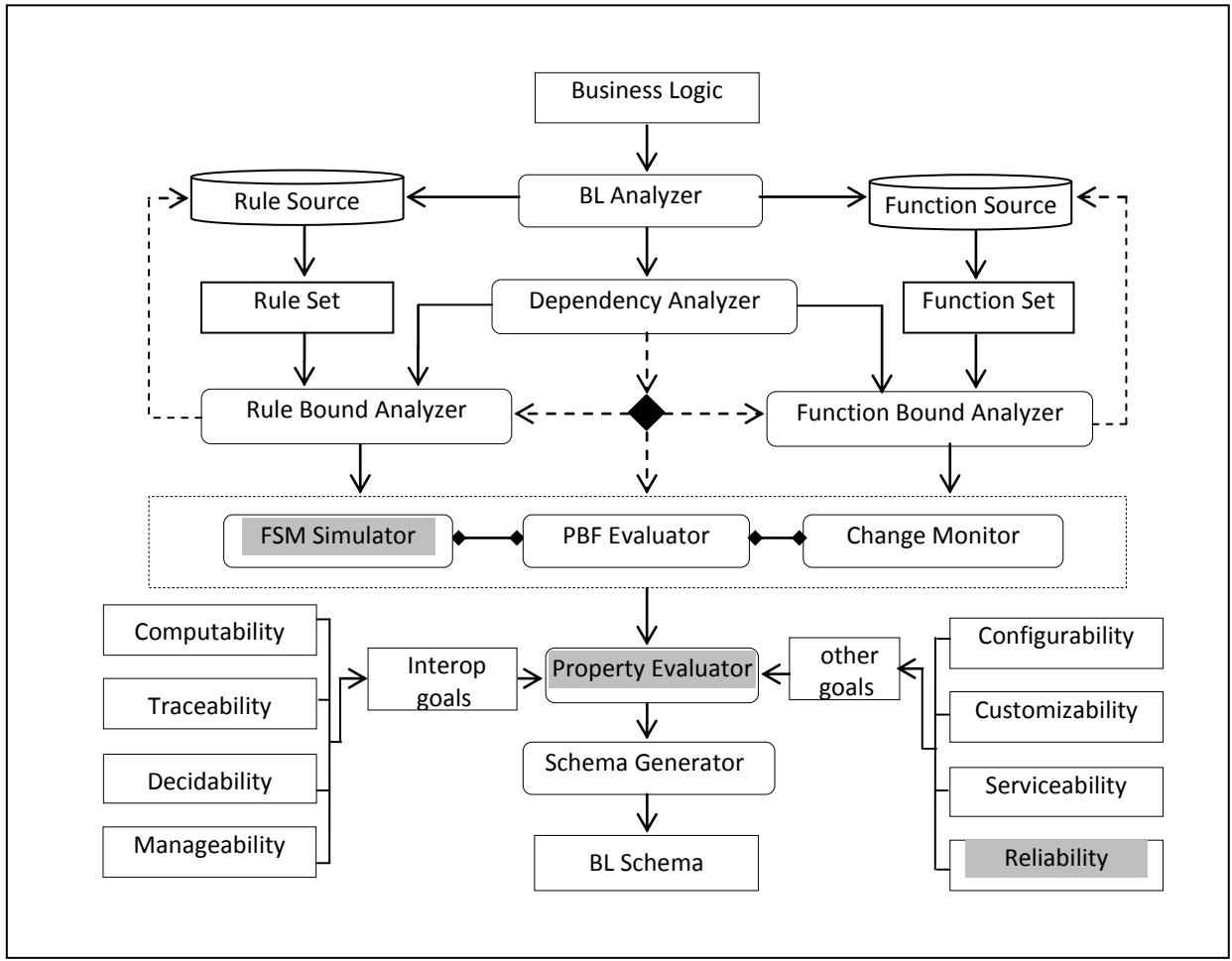

Figure1. Business Logic Evaluation Model

Business Logic Evaluation Model with a Business Logic Analyzer dissects the business logic into the rule source and function source that constitutes the logic. The rule source bears a collection of rules. The rule sets from rule source are analyzed by the rule bound analyzer. The function source contains all the functions of the business logic. The function sets from function source are analyzed by the function bound analyzer. The finite state machine simulator evaluates certain properties like computability and traceability whereas the PBF (Primitive Business Function) valuates properties such as configurability and customizability. The change monitor takes charge of Manageability and decidability. The property evaluator appraises the Business Logic with the interoperable goals such as Computability, Traceability, Decidability and Manageability and other goals such as Configurability, Customizability, Serviceability and Extensibility. After being done with the property evaluation the BL schema is generated from the schema generator.

## 3.1 Computability

Computability refers to feasibility of the modification anticipated. Whenever the request is issued, it is being edited by the analyst as a rule in order to bring in action the change request using the rule editor. If the request can be edited as rule using the editor, it means it is computable. And hence computability of a request is determined to be true or false based on whether it can be formed as rule by the editor or not. Therefore rule editor should consider all intricacies of the business product in its construct to effectively able to represent all computable rules. Thus the functional QoS parameter, computability is identified. It is enhanced by the construction of a functionally complete rule editor.

There are also other kinds of computability's defined, partial and complete computability. Say a rule includes a set of actions. If all the requirements of the action are satisfied it is said to be completely computable. If few of the requirements of the action are satisfied it is said to be partially computable. Say an action is composed of a data retrieval statement. If the specified database, table, fields are available, it is said to be computable. So in the editor we have also the space to specify actions constituted by a rule and we can calculate computability to provide more insight into computability of the rule.

## 3.2 Traceability

Rule traceability refers to tracing of similar rules which were solved before and issue of using its execution plan in solving current issue. Some errors may occur during runtime. The complaint is given by the user in general English and tries to edit it as rule using editor. In case of the same error occurring at two different locations, he may give the complaint in different way. If we manage to store the thread control block of that session then it is easy to identify similar change requests whose origin is due to the same internal operating state of the service. If the request happens to match and the issue was solved, we flag the present request as repeating request and indicate that the issue was not solved completely. If it happens to match partially, say parameters alone, we can make use of execution plan of the change request solved before.

## 3.3 Decidability

A rule or logic is said to be decidable if they are completely computable.( Each business logic system has both syntactic component which determines the notion of provability and the semantic component which determines the notion of logical validity). A rule or logic which is partially computable is sometimes said to be semi decidable. If we ascertain the rule is completely computable or partially computable we can determine the rule is decidable or semi decidable. This property is particularly useful in program verification, in the verification of reactive, real time or hybrid systems, as well as in databases and ontology. It is therefore important to identify such decidable fragments and design efficient decision procedures for them.

## 3.4 Manageability

Manageability is defined as a set of capabilities for discovering the existence, availability, performance, and usage, as well as the control and configuration of business logic within the Web services architecture. This implies that business logic can be managed using Web services technologies.

## 3.5 Configurability

The Configuration provides the functional capability to manage the collection of properties whose values may influence the behavior of a resource. Such deterministic behavior includes cost in terms of code size of business logic, slower algorithms, and so on. Hence these issues have to be managed to have a better configurability.

### 3.6 Customizability

Customizable solutions may be appropriate for customers whose needs and expectations change from time to time. Here the business analyst is given with customizing options so that he can change the logic according to the user's need.

### 3.7 Serviceability

Serviceability refers to the ability to identify exceptions or faults, debug or isolate faults to root cause analysis, and provide software maintenance in pursuit of solving a problem and restoring the problem in the logic. Incorporating serviceability facilitating features typically results in more efficient maintenance and reduces operational costs and maintains business continuity.

### 3.8 Reliability

#### 3.8.1 Reliability Analysis

Numerous web services are evolving today. Selecting the quality web service is a tedious process and one of the predominant QoS factor is reliability. Reliability is the ability of a system to keep operating over time. Software reliability is "the probability that the software will be functioning without failures under a given environmental condition during a specified period of time". A reliability evaluation framework model is designed. The Reliability depends on user profile. A module in a web service is defined to perform a particular function .Non executed part of codes have no influence on output. Little used modules might be less important for reliability of the system. Reliability of a module is the probability that the module performs the function correctly. A module passes result to next module.It is logical independent for design, programming and testing.

#### 3.8.2 Availability

Determine whether your users need reliability [non-performance has the greatest impact] or availability [downtime has the greatest impact] or both

$$A(t) = \frac{1}{1+t_m \lambda_F} \quad \text{or} \quad \lambda_F = \frac{1-A(t)}{t_m A(t)}$$

A = availability; tm = average downtime per failure

#### 3.8.3 Reliability

If you need to . . . convert from FIO to reliability

$$R = e^{-\lambda t}$$

– where R is reliability, λ = failure intensity, and t = number of natural or time units

If you need to . . . convert from reliability to FIO

$$\lambda = \frac{-\ln(R)}{t}$$

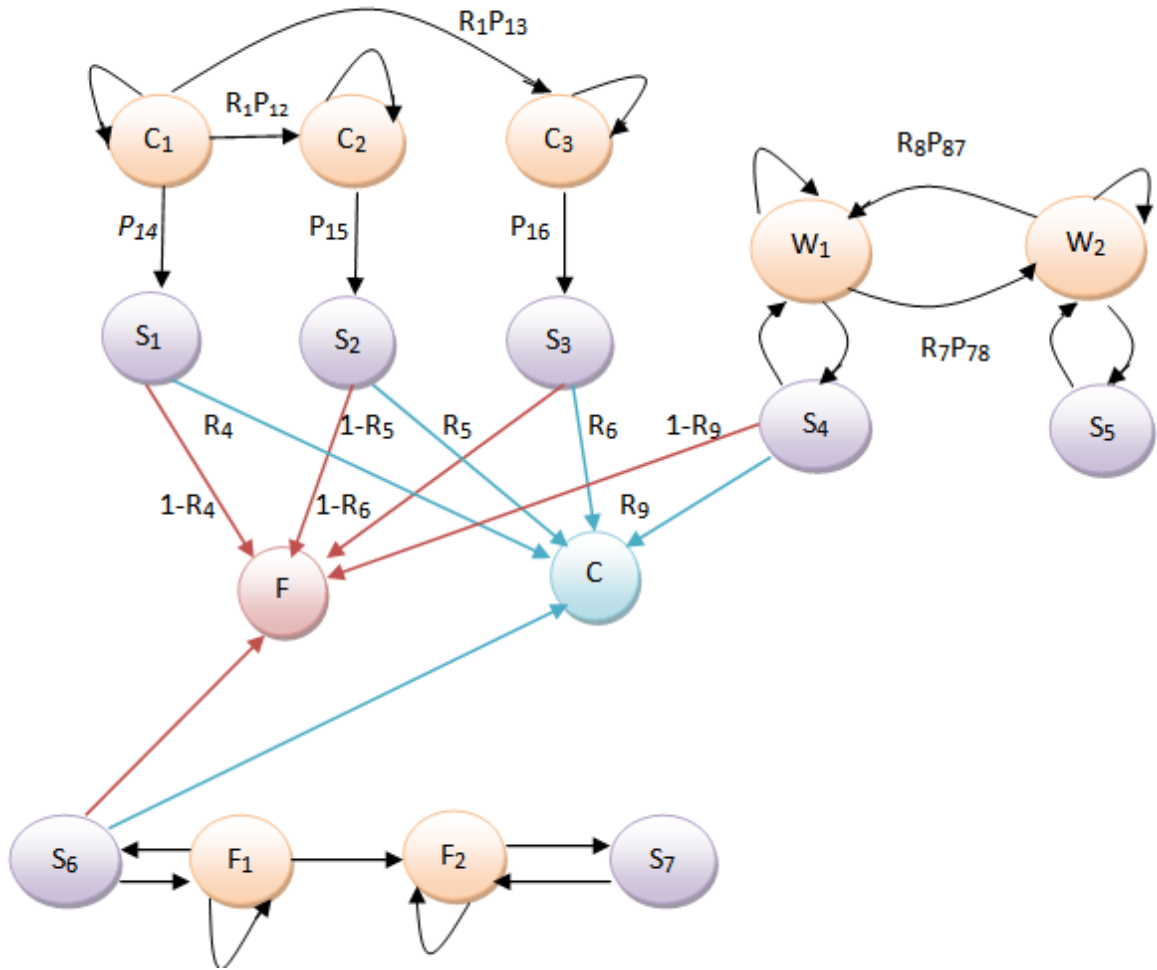

Figure2. Reliability Evaluation Model

In this model the conditional states are represented as nodes and each transition is given by a reliability factor R and probability factor P. Reliability factor is a probability that gets to the state C (state that returns correct output), when starting from any node say n. Probability factor values are decided based on the probability that the node performs function correctly and it varies between 0 and 1. In any service if majority of statements are directing to failure states i.e. F that service will be less reliable. Program runs correct only if the correct sequence of functions are executed and every function executed gives the correct result i.e C. Any service is said to be more reliable when modules of that service terminates in correct output C. Let us take an example of a program and evaluate its reliability using this model.

```
#include<stdio.h>
void main()
{
	int b,p,q,r,x;
	printf("enter the no: of rows : ");
	scanf("%d",&r);
	printf("\n pascals triangle :\n\n");
	b=1;
	q=0;
```

```
while(q<r)
        {
        for(p=30-3*q;p>0;p--)
                printf(" ");
        for(x=0;x<=q;x++)
                {
                if(x==0||q==0)
                        b=1;
                else
                        b=(b*(q-x+1)/x);
                printf("%6d",b);
                }
        printf("\n");
        q++;
        }
getch();
}
```

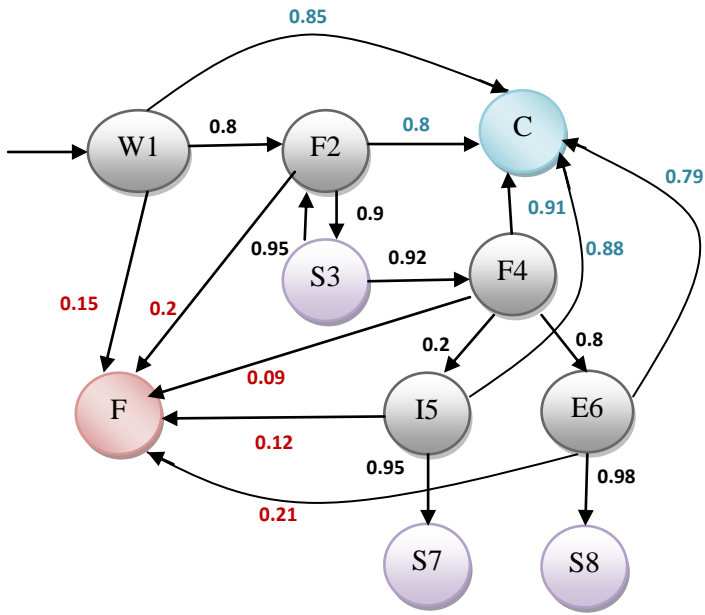

W1-while loop

F2- first for loop

S3- statements following F2

F4-second for loop

I5-if statement

E6-else statement

S7-statement following I5

S8-statement following E6

C, F-exit states

C-program returns correct output

F-program has a fault

Figure3. Reliability Evolution Model for the above program

The reliability evaluation model for this program is given in the above figure. The nodes represent the loops such as while, for followed by if – else statement. The sequence of the program is depicted in the above graph. Their transition represents probability that they will enter the next node. To the node F are the reliability factors (1-$R_i$) that the loop may have fault and to the node C are the reliability factors of correctness of the output. This program is reliable since the reliability factors of correctness sum to be higher than that of the fault. Thus these factors prove the program to be highly reliable.

# 4. WEB SERVICE AVAILABILITY EVALUATION USING SERVICE COMPOSITION

Availability of a web service over its life time during the web service composition is typically measured as a factor of its reliability - as reliability increases, so does availability. However, in service composition no composition method with its associated service composition set can guarantee 100.000% reliability; and as such, no active service composition set can assure 100.000% availability. Further, reliability in service composition involves a fine tuned selection process designed to optimize availability under a set of constraints, such as time and cost-effectiveness, activeness and compliance. Availability is the goal of most service users, and reliability evaluation and maintainability provide the means to assure that availability performance requirements are achieved.

## 4.1 Evaluation metrics

The most simple representation for **web service availability** is as a ratio of the expected value of the uptime of a service to the aggregate of the expected values of up and down time, or

$$WS_{Avail} = \frac{\delta[Service\_Uptime]}{\delta[Service\_Uptime] + \delta[Service\_Downtime]}$$

Let 'n' be the no of services participated in composition.

The Service composition set can be represented as $WS_{Comp} = \{S_1, S_2, \ldots S_n\}$

Availability for 'n' participating services is computed using the formula,

$$\sum_{i=1..n} WS_{Avail} = \sum_{i=1..n} \left[ \frac{\delta[Service\_Uptime]}{\delta[Service\_Uptime] + \delta[Service\_Downtime]} \right]$$

If we define the service monitoring function $M(t)$ as

$$M(t) = \begin{cases} 1, & \text{service continual functions at time 't'} \\ 0, & \text{otherwise} \end{cases}$$

Therefore, the service availability $WS_{Avail}(t)$ at time $t > 0$ is represented by

$$WS_{Avail}(t) = Pr[M(t)=1] = \delta[M(t)].$$

Average availability must be defined on an interval of the real line. If we consider an arbitrary constant $c > 0$, then average service availability is represented as

$$Avg[WS_{Avail}]_c = \frac{1}{c} \int_0^c WSAvail(t)\,dt.$$

Limiting (or steady-state) availability is represented by

$$WS_{Avail} = \lim_{t \to \infty} WSAvai(t)$$

Limiting average availability is also defined on an interval $(0,c]$ as

$$\text{Avg}[\text{WS}_{\text{Avail}}]_\infty = \lim_{c \to \infty} \text{WSAvai}_c = \lim_{c \to \infty} \frac{1}{c} \int_0^c \text{WSAvai}(t)dt, \quad c > 0.$$

For example, If a general purpose service is functioning which has mean time between failure (MTBF) of 8.15 years and mean time to recovery (MTTR) of 1 hour:

MTBF in hours = 8.15*365*24=71394

Availability= MTBF/(MTBF+MTTR) = 71394/71395 =99.9985%

Unavailability = 0.000141%

Table 1.Reliability evaluation for Web Service Composition

| Web Service Composition Set | | S_MTBF | S_MTTR | Availability |
|---|---|---|---|---|
| Travel Service | Reservation | 36441.6 | 36442.6 | 99.9972 |
| | Accommodation | 42924 | 42925 | 99.9976 |
| | Hotel | 41172 | 41173 | 99.9975 |
| Bank Service | Investment | 43800 | 43801 | 99.9977 |
| | Loan | 36792 | 36793 | 99.9973 |
| | Finance | 39420 | 39421 | 99.9974 |
| Search Service | Advanced Search | 40296 | 40297 | 99.9975 |
| | Quick Search | 38544 | 38545 | 99.9974 |
| | Keyword based Search | 42486 | 42487 | 99.9976 |

**Algorithm Reliability_Evaluation_WS_Composition**(WScomposition_Set,S_MTBF,S_MTTR)

---

**Input :** S_MTTR (Service mean time to recovery (MTTR) during service composition
        S_MTBF: Service mean time between failure (MTBF) during service composition
        S_Opr_Profile: Service Operational Profile
**Output:** WS_Avail : Web service availability over service composition
**Begin**
    S[]= Get WSCompositionSet 'S'
    // Web Service Composition Set
    $WS\_{comp}= \{S_1,S_2,…S_n\}$ // 'n' be the no of services
    For each service 'i' in 'S[n]'
    loop
    Find_in_S_Opr_Profile( $\delta[Service\_Uptime]$ )
    Find_in_S_Opr_Profile( $\delta[Service_{Downtime}]$ )

$$WS_{Avail} = \frac{\delta[Service\_Uptime]}{\delta[Service\_Uptime] + \delta[Service\_Downtime]}$$

$$WS_{Avail}[i] = WS_{Avail}$$

$$\Sigma_{i=1..n} WS_{Avail} = \Sigma_{i=1..n} \left[ \frac{\delta[Service\_Uptime]}{\delta[Service\_Uptime] + \delta[Service\_Downtime]} \right]$$

    Loop end
    // Find Average Service availability
    $WS_{Avail}(t)$ at time $t>0$ is represented by
    $WS_{Avail}(t) = Pr[M(t)=1] = \delta[M(t)]$.
    // consider an arbitrary constant $c > 0$, then average service availability is
    $Avg[WS_{Avail}]_c = \frac{1}{c} \int_0^c WSAvail(t)dt$.
**End**

## 5. CONCLUSIONS

In this paper we addressed actual need of reliability for an efficient web service. The web service and QOS grows expensively based on the reliability of the service. Hence we have proposed business logic evaluation model for web service reliability analysis using Finite State Machine (FSM) where FSM will be extended to analyze the reliability of composed set of service i.e., services under composition, by analyzing reliability of each participating service of composition with its functional work flow process. FSM is exploited to measure the quality parameters. If any change occurs in the business logic the FSM will automatically measure the reliability. we have also proposed an algorithm for reliability evaluation for web service composition which serves as to find which web service is available over service composition and finally we have evaluated the reliability for web service composition.